\def\year{2019}\relax
\documentclass[letterpaper]{article} 
\usepackage{aaai19}  
\usepackage{times}  
\usepackage{helvet} 
\usepackage{courier}  
\usepackage[hyphens]{url}  
\usepackage{graphicx} 
\usepackage{amsmath}
\usepackage{amssymb}
\usepackage{tikz}
\usepackage{multirow}
\usepackage[]{algorithm2e}

\def\MdR{\ensuremath{\mathbb{R}}}

\newcommand{\csch}[1]{{\color{blue}[CS: #1]}}
\renewcommand{\csch}[1]{}
\newcommand{\citet}[1]{\citeauthor{#1}~\shortcite{#1}}
\newcommand{\citep}{\cite}

\newcommand{\ie}{i.e.\ }

\newcommand{\Is}       {:=}
\newcommand{\sodass}{\,:\,}
\newcommand{\setGilt}[2]{\left\{ #1\sodass #2\right\}}
\newcommand{\set}[1]{\left\{ #1\right\}}
\newcommand{\gilt}{:}

\usepackage{graphicx}  
\nocopyright
\title{Finding Optimal Longest Paths by Dynamic Programming in Parallel}
\author{Kai Fieger\textsuperscript{\rm 1}, Tom\'a\v{s} Balyo\textsuperscript{\rm 1}, Christian Schulz\textsuperscript{\rm 2}, Dominik Schreiber\textsuperscript{\rm 1}\\ 
\textsuperscript{\rm 1}Karlsruhe Institute of Technology\\ 
Karlsruhe, Germany\\
\textsuperscript{\rm 2}University of Vienna, Faculty of Computer Science\\
Vienna, Austria\\
fieger@ira.uka.de, tomas.balyo@kit.edu,\\
christian.schulz@univie.ac.at, dominik.schreiber@kit.edu 
}
 
\begin{document}

\maketitle

\begin{abstract}
We propose an exact algorithm for solving the longest simple path problem between
two given vertices in undirected weighted graphs. 
By using graph partitioning and dynamic programming, we obtain an algorithm 
that is significantly faster than other state-of-the-art methods. This enables us to solve instances that were previously unsolved and solve hard instances significantly faster.
Lastly, we present a scalable parallelization which yields the first efficient parallel algorithm for the problem.
\end{abstract}

\section{Introduction} 
The longest path problem (LP) is to find a simple path (each vertex visited at most once)
of maximum length between two given vertices of a graph, 
where length is defined as the number of edges or the total weight of the edges in the path.
The problem is known to be NP-complete~\cite{NP} and has several applications such as designing circuit boards~\cite{Circuit1,Circuit2}, project planning~\cite{Brucker}, information retrieval~\cite{Wong} or patrolling algorithms for multiple robots in graphs~\cite{Multirobot}. 

In this paper we present the algorithm LPDP (Longest Path by Dynamic Programming) and its parallel version. LPDP makes use of graph partitioning and dynamic programming. Unlike many other approaches for NP-complete problems, LPDP is not an approximation algorithm -- it finds an optimal longest path.
Through experiments we compare LPDP to previous LP algorithms and evaluate the speedups achieved by the parallel algorithm.

\section{Preliminaries}
\label{s:basicconcepts}
In the following we consider an \emph{undirected} graph $G=(V,E,\omega)$ 
with (symmetric) edge weights $\omega: E \to \MdR_{\geq 0}$, $n = |V|$, and $m = |E|$.
We extend $\omega$ to sets, i.e.,
$\omega(E')\Is \sum_{e\in E'}\omega(e)$.
$N(v)\Is \setGilt{u}{\set{v,u}\in E}$ denotes the neighbors of $v$.
A \emph{subgraph} is a graph whose vertex and edge sets are subsets of another graph. We call a subgraph \emph{induced} if every edge among the included vertices is included.   
A subset of a graph's vertices is called a \emph{clique} if the graph contains an edge between every two distinct vertices of the subset.
A \emph{matching} is a subset of the edges of a graph where no two edges have any vertex in common.
A sequence of vertices $s \to \cdots \to t$ such that each pair of consecutive vertices is connected by an edge, is called an \emph{$s$-$t$ path}. 
We say that $s$ is the source and $t$ is the target. 
A path is called \emph{simple} if it does not contain a vertex more than once.
The length of a path is defined by the sum of its edge weights. 
If the graph is unweighted, then edge weights are assumed to be one.

Given a graph $G=(V,E,\omega)$ as well as two vertices $s, t \in V$, the \emph{longest path} (LP) problem is to find the longest simple path from $s$ to $t$. 
Another version of the LP problem is to find the overall longest simple path in the graph. 
However, the problem can be solved by introducing two vertices $s, t$, connecting them to all other vertices in the graph by edges of weight zero and then running algorithms tackling the LP problem
on the modified instance.

A $k$-way partition of a graph is a division of $V$ into \emph{blocks} of vertices $B_1$,\ldots,$B_k$, \ie $B_1\cup\cdots\cup B_k=V$ and $B_i\cap B_j=\emptyset$
for $i\neq j$.
A \emph{balancing constraint} demands that 
$\forall i\in \{1..k\}\gilt |B_i|\leq L_{\max} := (1+\epsilon)\lceil\frac{|V|}{k}\rceil$
for some imbalance parameter $\epsilon$. 
The objective is typically to minimize the total \emph{cut} $\sum_{i<j}\omega(E_{ij})$ where 
$E_{ij}\Is\setGilt{\set{u,v}\in E}{u\in B_i,v\in B_j}$.

\subsection{Related Work}
Previous work by \citet{stern} mainly 
focuses on the possibility of applying algorithms that are usually used to 
solve the shortest path problem (SP) to the longest path problem. 
\citet{stern} make clear why LP is so difficult compared to SP. 
Several algorithms are presented that are frequently used to solve SP or other minimization search problems.
They are modified in order to be able to solve LP. 
The search algorithms are put into three categories: heuristic, uninformed and suboptimal. 
Each of the algorithms in the first two categories yields optimal solutions to the problem.
The most relevant category for this paper is heuristic searches. Here, a heuristic can provide 
extra information about the graph or the type of graph. Heuristic searches 
 require a heuristic function that estimates the remaining 
length of a solution from a given vertex of the graph. 
This can give important information helping to prune the search space and to speed up the search.
\citet{stern} show that heuristic searches can be used efficiently for the longest path problem. 
Some examples of algorithms in this category are Depth-First-Branch-and-Bound (DFBnB) and~A*.
Another category represents ``uninformed'' searches, which do not require any information other 
than what is already given in the definition of the problem. 
Examples from this category are Dijkstra's algorithm or DFBnB without a heuristic. Modifying 
these algorithms to fit LP basically leads to brute force algorithms, which means 
that they still have to look at every possible path in the search space. 
Hence, these uninformed search strategies are not very beneficial for LP. 
The last category are suboptimal searches. The authors looked at a large 
number of these algorithms that only find approximations of a longest path. 

A similar problem to LP called \emph{Snake in the Box} (SIB) is the problem of finding the
longest simple path in an $n$-dimensional hypercube.
Additionally to the constraint of not allowing repeated
vertices in the path it is required that for any two vertices $u,v$ there is no edge between $u$ and $v$
unless $u$ and $v$ are adjacent in the path. Heuristic search LP algorithms can be adapted to efficiently
solve SIB~\cite{palombo2015solving} by designing SIB specific heuristics. However, these techniques cannot be transferred to solving the general LP problem since they rely on SIB specific heuristics and therefore these results are not relevant for~our~work.

The LP problem is related to the Traveling Salesperson Problem (TSP), which is a very well studied
problem. There exist several exact algorithms and solvers for TSP, such as the Integer Linear Programming based exact  solver Concorde~\cite{concorde}. 
TSP problems can be solved by translating them into LP problems and using an LP 
solver~\cite{hardgrave1962relation}. The translation is very efficient since it only adds one new vertex
and at most $n$ new edges, where $n$ is the number of vertices of the TSP problem. This raises the
question if we can solve TSP problems faster by using our new algorithm (after translating them to LP) than the state of the art TSP solver Concorde~\cite{concorde}. If we consider the standard TSP benchmark 
problems, i.e., the TSPLib collection~\cite{tsplib}, the answer is no. The reason is that all the TSPLib
benchmark problems are cliques and they remain cliques after translating them to LP~\cite{hardgrave1962relation}. This is very unfortunate, since our algorithm relies on
graph partitioning, which is not very helpful for cliques.
Perhaps for graphs that are not cliques and can be well partitioned our algorithm could outperform Concorde.
On the other hand, translating LP to TSP is also 
possible~\cite{lawler1985traveling}. Nevertheless, this translation introduces a lot of auxiliary vertices and
edges. Indeed, the number of vertices increases by a factor of 6 and the number of edges is quadratic in the
number of vertices (both original and auxiliary). This means that even problems that are solved in milliseconds
using our new algorithm become unsolvable
by Concorde after translating them to TSP (according to our experiments).
In summary, we conclude that although TSP and LP can be reduced to each other it is best to solve each
problem with its own dedicated solver.

\section{Longest Path by Dynamic Programming}
\label{s:maincontribution}
\label{sec:algo}
We now introduce the main contribution of our paper which is a new algorithm to tackle the longest path problem based on principles of dynamic programming. 
Hence, our algorithm is called ``Longest Path by Dynamic Programming'' (LPDP). 
Our algorithm solves the longest path problem (LP) for \emph{weighted undirected graphs}. 

\subsection{Exhaustive Depth First Search}
\label{sec:approach}
A simple way to solve the longest path problem is \emph{exhaustive depth-first search}~\cite{stern}. 
In regular depth-first search (DFS) a vertex has two states: marked and unmarked. Initially, all vertices are unmarked. The search starts by calling the DFS procedure with a given vertex as a parameter. This
vertex is called the root.
The current vertex (the parameter of the current DFS call) is marked and then the DFS procedure is recursively executed on each unmarked vertex reachable by an edge from the current vertex. The current vertex is called the parent of these vertices. Once the recursive DFS calls are finished we backtrack to the parent vertex. The search is finished once DFS backtracks from the root.

\begin{algorithm}
\SetKwProg{Fn}{Search}{}{}
\Fn{exhaustiveDFS($v$)}{
\If{$v$ is unmarked}{
   mark $v$\;
   \ForEach{$ \{v,w\} \in E$}{
   	exhaustiveDFS($w$)\;
   }
   unmark $v$\;
  }
}
 \caption{Exhaustive depth first search. In order to solve LP we start this search from the start node and update the best found solution each time the (unmarked) target node is found.}
\label{exDFS}
\end{algorithm}

Exhaustive DFS is a DFS that unmarks a vertex upon backtracking. In that way every simple path 
in the graph starting from the root vertex is explored. The LP problem can be solved with 
exhaustive DFS by using the start vertex as root. During the search the 
length of the current path is stored and compared to the previous best 
solution each time the target vertex is reached. If the current length is greater than 
that of the best solution, it is updated accordingly. When the search is done 
a path with maximum length from $s$ to $t$ has been found. If we store the length of longest path for each vertex (not just the target vertex) then all the longest simple paths from $s$ to every other vertex can be computed simultaneously.

It is easy to see, that the space complexity of exhaustive DFS is the same as regular DFS -- linear in the
size of the graph. However, the time complexity is much worse. In the worst case -- for a clique with $n$
vertices -- the time
complexity is $\mathcal{O}(n!)$ since every possible simple path is explored, which corresponds to all
the permutations of the vertex set. If the maximum degree of the graph is $d$ then the running time can be bound
by $\mathcal{O}(d^{n})$, where $n$ is the number of vertices.

\subsection{Algorithm Overview}
\begin{figure}[t]
\centering
\begin{tabular}{cc}
\includegraphics[width=0.25\textwidth]{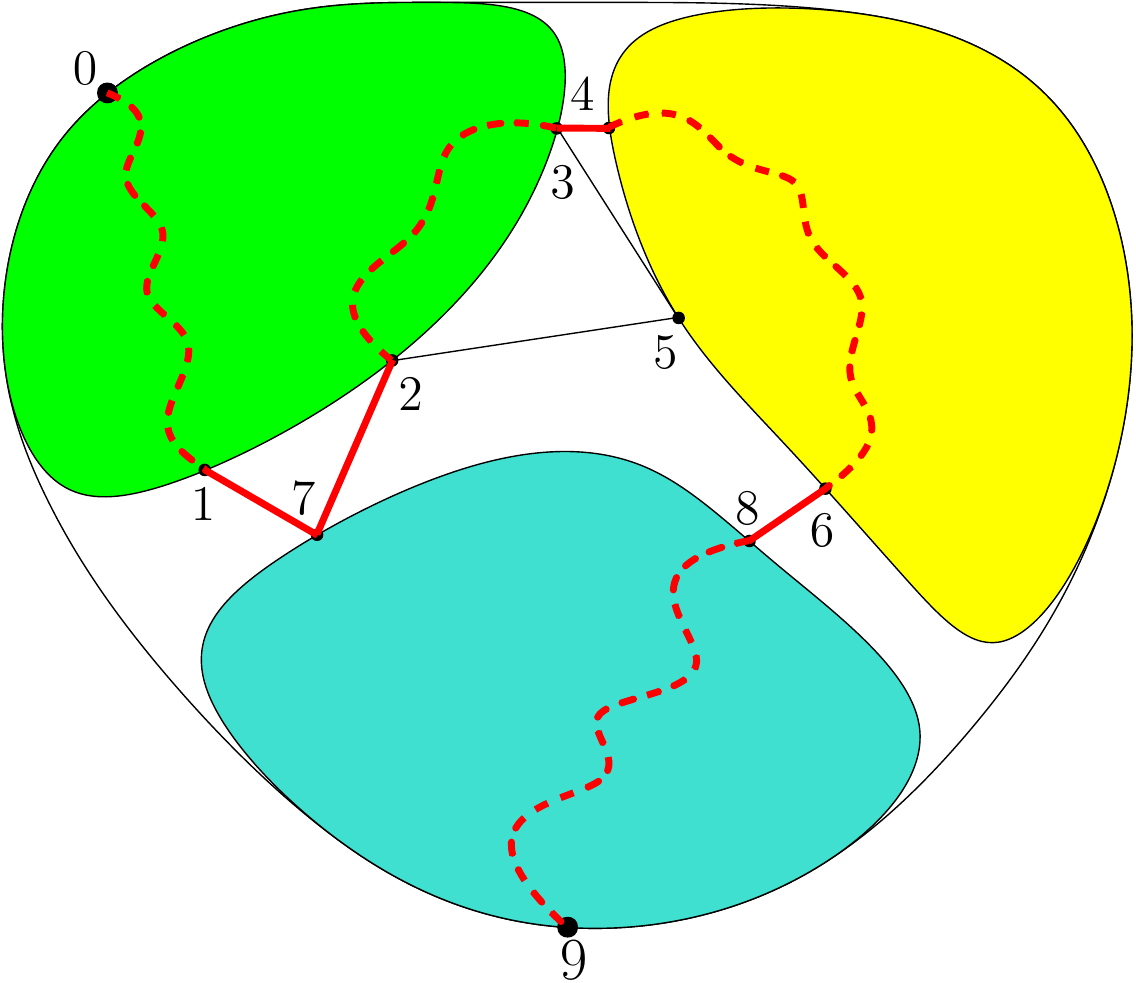}&\includegraphics[width=0.15\textwidth]{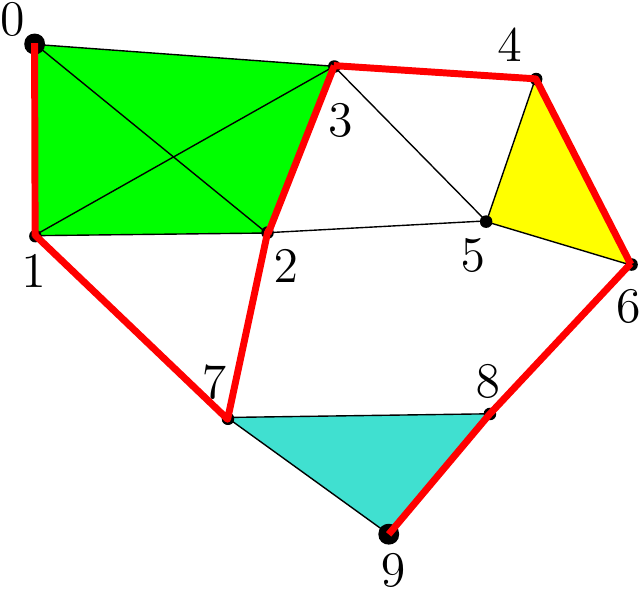}
\end{tabular}
\caption{Left shows an example of a graph that is partitioned into three blocks. The nodes 0 to 9 are the boundary nodes of the blocks. The edges between them are the boundary edges. A possible longest path from 0 to 9 is shown in red. The right graph is the auxiliary graph that is constructed by the LPDP algorithm. The path that corresponds to the one on the left is shown. The induced boundary node pairs for the blocks are: $P_{green} = \{\{0,1\}, \{2,3\}\}$, $P_{yellow} = \{\{4,6\}\}$, $P_{blue} = \{\{7,7\}, \{8,9\}\}$}
\label{fig:example}
\end{figure}

Our algorithm is based on dynamic programming: roughly speaking, we partition the graph into blocks, 
define subproblems based on the blocks and then combine the solutions into a longest path for the original problem.
In order to be able to divide LP into subproblems, we first generalize the problem: \\\textbf{Given a graph} $G=(V,E,\omega)$, 
two vertices $s,t \in V$, two sets $B \subseteq V$ and $P \subseteq \{\{u,v\} \mid u,v \in b(B) \}$ where $b(B) := \{ v \in B \mid v = s \lor  v = t \lor \exists{} \{v,w\} \in E : w \notin B \}$ 
are the $boundary$ $nodes$ of $B$, \textbf{find a simple path} from $a$ to $b$ in the 
subgraph induced by $B$ for every $\{a,b\} \in P$. 
Find these paths in such a way that they do not intersect and have the maximum possible cumulative weight.
See Figure~\ref{fig:example} for an example.\\

\noindent We make the following Observations about this problem:
\begin{enumerate}
    \item\label{structureOfP} A pair $\{a,a\} \in P$ is possible and results in a path of weight 0 that consists of one node and no edges. But otherwise the problem is unsolvable if any node occurs twice in $P$. This would result in two intersecting paths as they would have a common start or end node.
    \item\label{MX} We calculate an upper bound on the number of all solvable $P$ in the following way: We transform~$P$ into two sets $(M,X)$. $\{x,y\} \in P \land x \neq y \iff \{x,y\} \in M$ and $\{x,x\} \in P \iff x \in X$. We interpret $M$ as a set of edges in the clique of the boundary nodes $b(B)$. It follows from Observation \ref{structureOfP} that $M$ represents a matching (set of edges without common vertices) in that clique. The numbers of all possible matchings in a clique of size $n$ are also known as the telephone numbers or involution numbers \cite{involution}: $T(n) := \sum\limits_{k=0}^{\lfloor n / 2 \rfloor} \frac{n!}{2^{k}(n-2k)!k!}$. Each element of the sum equals the number of matchings with $k$ edges. Any of the $n-2k$ boundary nodes that are left are either in $X$ or not. This leads to $2^{n-2k}$ possibilities for $X$ per $k$-edge matching $M$. This means that there are at most $\sum\limits_{k=0}^{\lfloor n / 2 \rfloor} \frac{n!2^{n-3k}}{(n-2k)!k!}$ possible, solvable $P$. It is the exact number if the subgraph induced by $B$ is a clique.
    \item\label{unsolvable} If the problem is unsolvable for a $P$ it is also unsolvable for any $P' \supseteq P$. 
    \item\label{subproblem} A solution of the problem also induces a solution to the problem for any $B' \subseteq B$ and a $P'$: Restricting the solution to nodes of $B'$ results in non-intersecting, simple paths in the subgraph of $B'$. These paths start and end in boundary nodes of $B'$ inducing the set $P'$.\\
    These paths are the solution to the problem for $B'$ and $P'$. \emph{Proof:} Otherwise we could start with the solution for $B$ and $P$, remove all paths in the subgraph of $B'$ and replace them with the solution for $B'$ and $P'$. We would obtain a solution for $B$ and $P$ with a higher cumulative weight than before. This is impossible.
\end{enumerate}

LP is a special case of this problem where $B = V$  and $P = \{\{s,t\}\}$. Observation \ref{subproblem} is the basis of the LPDP algorithm as it allows us to recursively divide LP into subproblems. 
LPDP requires a hierarchical partitioning of the graph. Level 0 represents the finest level of partitioning. On each higher level we combine a group of blocks from the lower level into a single larger block. On the highest level we are left with a single block $B = V$. We solve our problem for each of these blocks and any possible $P$: We start by calculating the solutions for each block of level~0. We then calculate the solutions for a block on level 1 by combining the solutions of its level 0 sub-blocks. This is repeated level by level until we calculated all solutions for the block $B = V$, namely the solutions for $P= \{\{s,t\}\}, \{\}, \{\{s,s\}\}, \{\{t,t\}\}$ and $\{\{s,s\}, \{t,t\}\}$. The latter four are trivial (see Observation \ref{structureOfP}) and do not have to be calculated. With a solution for $P= \{\{s,t\}\}$ we solve LP.

The next section shows how we calculate the solutions for one block $B$ with the help of its sub-blocks from the level below. The initial solutions for the blocks on level 0 can be calculated with the same algorithm. In order to do this we interpret each node $v$ as a separate sub-block. We know the solutions for each of these sub-blocks ($P= \{\}~or~\{\{v,v\}\}$). So we can use the same algorithm to calculate solutions for the blocks on level 0.

\subsection{Combining Solutions}
\label{sec:combsolutions}
Let $P_S$ be the set of boundary node pairs for any set of nodes $S$. Given is the subset of nodes $B$ of the graph and a partition $B_1$,\ldots,$B_k$ of $B$ ($B_1\cup\cdots\cup B_k=B$ and $B_i\cap B_j=\emptyset$ for $i\neq j$). We already know the solution of the problem for each $B_i$ and every possible $P_{B_i}$. We calculate the solution for $B$ and every possible $P_B$ with the following algorithm:\\
We construct an auxiliary graph $G' = (V', E', \omega')$ with $V' = \bigcup\limits_{i=1}^{k} b(B_{i})$. $E'$ contains all edges $\{v,w\} \in E$ where $v \in b(B_i)$ and $w \in b(B_j)$ (with $i \neq j$). We call these edges \emph{boundary edges}. They keep the weight they had in $G$. We also create a clique out of the boundary nodes of every $B_i$. These new edges have zero weight. 

In order to calculate the solutions for $B$, we start a modified version of the exhaustive DFS on every boundary node of $B$. Pseudocode of this search algorithm is shown in Algorithm \ref{lpdpsearch}. Compared to exhDFS it works with multiple paths. The first path starts from the starting boundary node. Once another boundary node of $B$ is reached, the current path can be completed. Then search starts a new path from another boundary node. At any point of the search $P_B$ is equivalent to the boundary node pairs induced by the completed paths. The sets $P_{B_i}$ are maintained the following way: The paths contain an edge $\{v,w\}$ of the $B_i$-clique $\iff \{v,w\} \in P_{B_i}$. If the paths contain a node $v \in B_i$ but no edge $\{v,w\}$ of the $B_i$-clique: $\{v,v\} \in P_{B_i}$. During the search we do not traverse an edge that would induce a $P_{B_i}$ without a solution. Further traversal of a path with an unsolvable $P_{B_i}$ only leads to $P'_{B_i} \supseteq P_{B_i}$ which is still unsolvable (as already seen in Observation \ref{unsolvable}).

Each time we complete a path, we have calculated a candidate for the solution to $B$ and $P_B$. The weight of this candidate is the weight of the solution of each block $B_i$ and the induced $P_{B_i}$ plus the weight of all boundary edges in the paths. Until now, no $P_B$ found by the search contains a pair $\{v,v\}$ as we do not allow a path to end in its starting boundary node. This way $P_B$ is equivalent to a $M$ and $X = \emptyset$ according to the representation in the Observation \ref{MX}. So when we complete a path, we additionally go through all possible sets $X$ (while modifying the sets $P_{B_i}$ accordingly) and update the best found solution for these candidates as well. This leads to a faster search in the auxiliary graph compared to letting a path end in its starting node.

An important optimization which can be seen in Algorithm \ref{lpdpsearch} is that we only allow a path to end in a boundary node with a higher id than its starting boundary node. Additionally a path can only start from a boundary node with a higher id than the starting node of the previous path. The first optimization essentially sorts the two vertices of each pair $\{x,y\} \in P$. The second then sorts these pairs. This results in an order and a direction in which we have to search each path in $P$. This avoids unnecessary traversal of the graph.

\begin{algorithm}
\SetKwProg{Fn}{Search}{}{}
\Fn{LPDP-Search($v$)}{
\If{$v$ unmarked $\&$ $\forall i\ \exists$ a solution for $B_i$ and  $P_{B_i}$}{
   mark $v$\;
   \If{$v \in b(B)$}{
	\eIf{already started a $\{a,\cdot\}$-path}{
		\If{$v > a$}{
			$\{a,v\} \in P_B$\;
			\ForEach{$w \in b(B)$ where $w > a$}{
   				LPDP-Search($w$)\;
   			}
			$\{a,v\} \notin P_B$\;
		}
   	}{
   		start a $\{v,\cdot\}$-path\;
   	}
   }
   \ForEach{$ \{v,w\} \in E$}{
   	LPDP-Search($w$)\;
   }
   unmark $v$\;
}
}
 \caption{Basic search algorithm that is used to search the auxiliary graphs.}
\label{lpdpsearch}
\end{algorithm}

In the worst case scenario each block on every level of the partition hierarchy is a clique. According to Observation~\ref{MX} this means we have to store  $\sum\limits_{k=1}^{\lfloor n / 2 \rfloor} \frac{n!2^{n-3k}}{(n-2k)!k!}$ solutions for each block with $n$ vertices. Note that we start the sum with $k=1$. The element of the sum with $k=0$ represents the number of all $P$ that do not contain any $\{x,y\}$ where $x \neq y$. These solutions always exist and have weight 0. We do not have to store them.

In our implementation we use a hash table for every block $B$ to store its solutions. The hash table uses $P_B$ as the key and stores the induced $P_{B_i}$ and the overall weight as a value. On the lowest level we store the paths instead. When the algorithm is finished we recursively unpack the longest path by looking up each $P_{B_i}$ in the hash table of its block.

\section{Parallelization}
\label{c:parallelization}
The parallelization of the LPDP algorithm is done in two ways. First, multiple blocks can be solved at the same time. Additionally, LPDP-Search from Algorithm \ref{lpdpsearch}, which is used to solve a block, can also be parallelized.

\subsection{Solving Multiple Blocks}
A block is only dependent on its sub-blocks. We can solve it once all of its sub-blocks have been solved. No information from any other block is needed. This allows us to solve multiple blocks independently of each other. 
The question is how effective this form of parallelism can be. For example: If p\% of a problem's serial runtime is spent solving a single block, solving multiple blocks in parallel cannot achieve a speedup higher than $\frac{100}{p}$. In order to test this we ran the serial LPDP solver on the set of problems that is used in the experiment section. LPDP had a time limit of 64 minutes to solve a problem. We only looked at problems that took more than 5 seconds to solve. 

\begin{figure}[]
  \includegraphics[width=\linewidth]{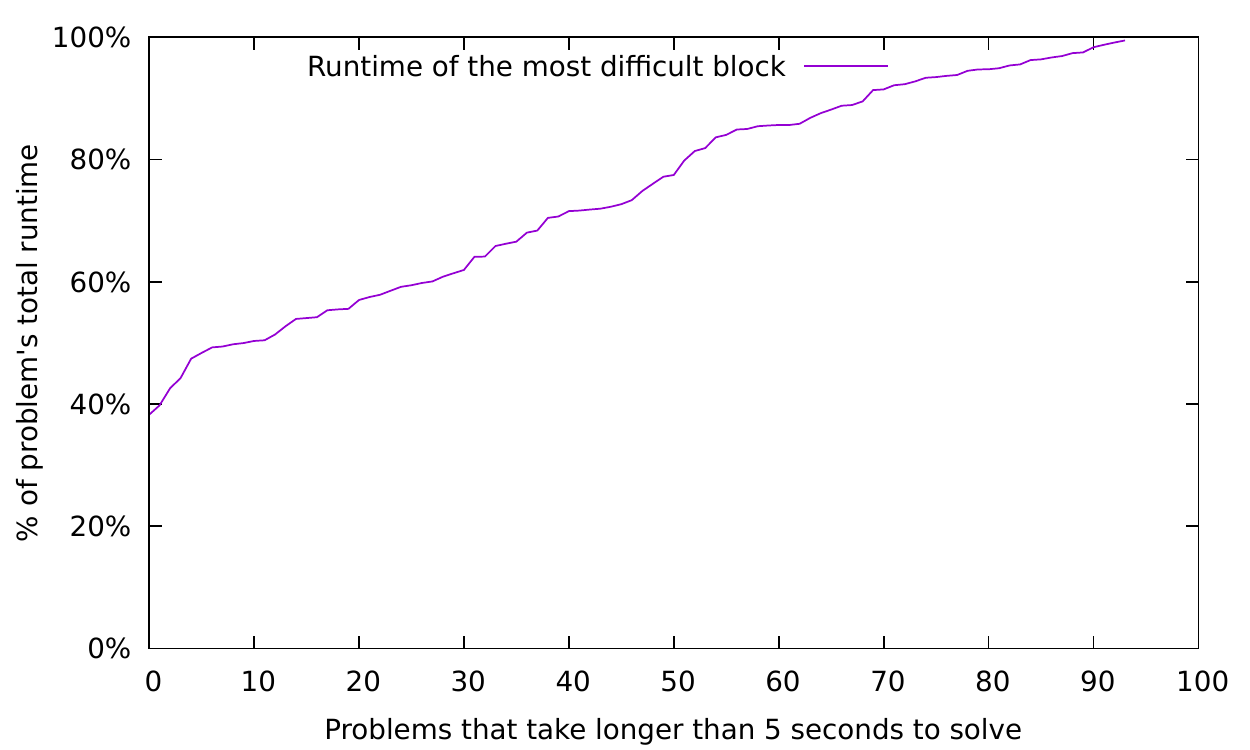}
  \caption{The plot is restricted to the problems that took the serial LPDP solver more than 5 seconds. For each of these problems the plot shows the percentage of the problem's runtime that is spent on solving its most difficult block. The problems are sorted from lowest to highest percentage.}
  \label{fig:single_block_percentage}
\end{figure}

For each of the solved problems the plot in Figure \ref{fig:single_block_percentage} shows the percentage of the problem's runtime that is spent solving its most difficult block. The problems are sorted from lowest to highest percentage. From this plot we can see that achieving speedups over 2 would be impossible for most of the problems.  In fact for almost half of the shown problems a single block makes up over 80\% of their runtime. This means that parallelizing the execution of a single block is far more important than solving multiple blocks at the same time. The next section explains how this is done.

\subsection{Parallelizing LPDP-Search}
\label{sec:parallel_lpdp_search}
In order to solve a block, LPDP starts the search shown in Algorithm \ref{lpdpsearch} from each boundary vertex of the block (except the last). We could parallelize the solving of a block simply by running multiple of these searches at the same time. There are multiple problems with this approach:
We could only use up to $n-1$ threads when solving a block with $n$ boundary vertices. This limits the speedup that could be achieved to $\frac{1}{n-1}$. Additionally, because of the optimization explained in Section \ref{sec:combsolutions}, the searches that start from boundary vertices with lower ids usually take much longer than those started from higher boundary vertices. This would lead to bad load balancing and limit the possible speedup even further.

\begin{figure}[]
  \centerline{\includegraphics[width=\linewidth]{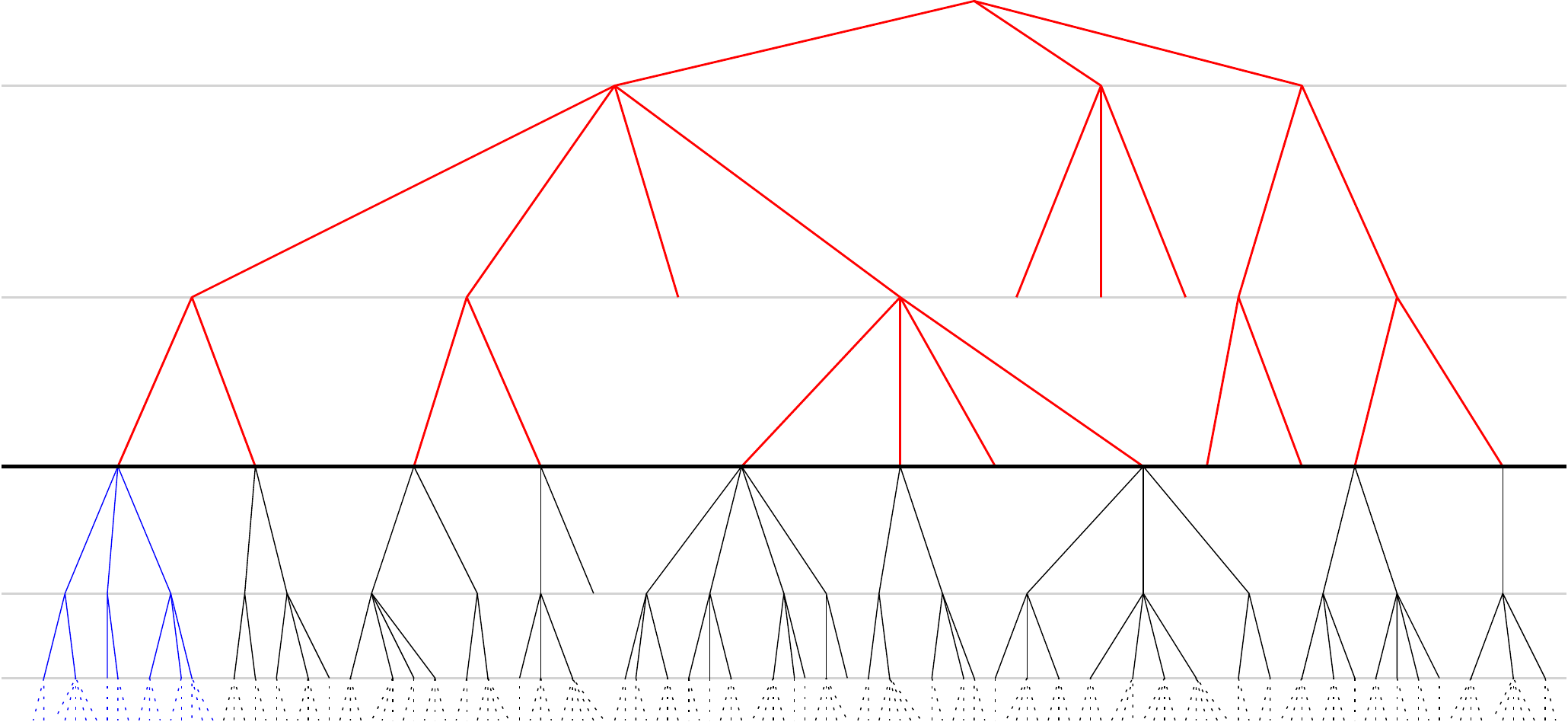}}
  \caption{This tree is an example that is generated through the recursion of an LPDP-Search. The horizontal lines are the different levels of recursion. The root of the tree is the initial call of LPDP-Search(). Every edge represents a function call to LPDP-Search(). For example the initial LPDP-search() calls itself 3 times. This results in the 3 edges to the first horizontal line (first level of recursion). The call represented by the left edge calls LPDP-Search() 4 times. This results in 4 branches to the second horizontal line. The whole tree represents a full LPDP-Search. In red we can see a LPDP-Search that is limited to 3 levels of recursion.}
  \label{fig:recursiontree}
 \end{figure}

An approach that does not have these problems is inspired by the ``Cube and Conquer'' approach for SAT-Solving that was presented by \citet{heule}. In this paper a SAT formula is partitioned into many subformulas. These subformulas can be solved in parallel. We can do the same for LPDP by partitioning the search space of LPDP-Search into many disjoint branches. We do this by running LPDP-Search from each boundary vertex with a limited recursion depth. Every time the search reaches a certain level of recursion the search stores its current context in a list and returns to the previous recursion level. Figure \ref{fig:recursiontree} shows an example of this. We can see a LPDP-Search limited to 3 recursions in red. A stored context represents all the data that allows us to continue the search at this point later on. On higher recursion levels the search works the same as before. The created list of contexts is then used as a queue. Each element of the queue represents a branch of the search that still has to be executed. One such branch can be seen in blue in Figure \ref{fig:recursiontree}.

We execute these branches in parallel. Each time a thread finishes one branch it receives the next branch from the top of the queue. This automatically results in a form of load balancing as threads that execute faster branches simply end up executing more branches. In order for this to work well we need a large number of branches. But generating the branches should also not take too long. We have to choose the recursion depth limit accordingly. This could be done by initially choosing a small recursion depth limit. If the resulting list of branches is too small, we repeat the process with a higher and higher limit until the list has the necessary size. These repeats should not take too long as a small list of search branches would also indicate a short runtime of the restricted search. If we want to prevent the repeats, we could also continue the restricted LPDP-Search from each of the branches in the list until a higher recursion depth is reached. In the experiments none of this was done. Instead the recursion depth limit was set to the fixed value 5. This has proven to be sufficient. Generating the branches only took a negligible amount of time but still resulted in good load balancing.

This form of parallelization requires a certain amount of synchronization between the threads. The serial implementation of LPDP uses hash tables to store the solutions for a block. Several threads can try to modify the same hash table entry at the same time. Introducing a concurrent hash table\footnote{In our implementation
we use the concurrent hash table from the TBB~\cite{DBLP:reference/parallel/Robison11} library.} solves this problem.

\section{Experimental Evaluation}
\label{s:experiments}
\paragraph*{Methodology.} We have implemented the algorithm described above using C++ and compiled it using gcc 4.9.4 with full optimizations turned on (\texttt{-O3} flag). 
Our implementation is freely available in the Karlsruhe Longest Paths package (KaLP) under the GNU~GPL~v2.0 license ~\cite{kaLPHomePage}.  
For partitioning the input graph we use the partitioner KaHIP~\cite{KaHIP}.
We use multiple implementations provided by \citet{stern} for comparison: $\textbf{Exhaustive DFS}$ is the naive brute-force approach as well as the $\textbf{A*}$ algorithm and the $\textbf{DFBnB}$ solver.
We run each algorithm and input pair with a time limit of one hour.
Experiments were run on a machine that is equipped with four Intel\textregistered{} Xeon\textregistered{} Processors E5-4670 (2.4 GHz with 8 cores each -- 32 cores in total) and 512 GB RAM. 

We present multiple kinds of data:
first and foremost, we use \emph{cactus plots} in which the number of problems is plotted against the running time. The plot shows the running time achieved by the algorithm on each problem. The running times are sorted in ascending order for each algorithm. The point ($x$, $t$) on a curve means that the $x$th fastest solved problem was solved in $t$ seconds. Problems that were not solved within the time limit are not shown. In addition we utilize tables reporting the number of solved problems as well as \emph{scatter plots} to compare running times of two different solvers $\mathcal{A},~\mathcal{B}$ by plotting points $(t_\mathcal{A}, t_\mathcal{B})$ for each instance.

\paragraph*{Benchmark Problems.}
We mainly use instances similar to the ones that have been used in previous work by \citet{stern}, \ie based on mazes in grids as well as the road network of New York. Additionally we use subgraphs of a word association graph~\cite{webgraphWS1,webgraphWS2}. The graph describes the results of an experiment of free word association performed by more than 6000 participants. Vertices correspond to words and arcs represent a cue-target~pair.

\begin{figure}[t!]
  \centerline{\includegraphics[width=\linewidth]{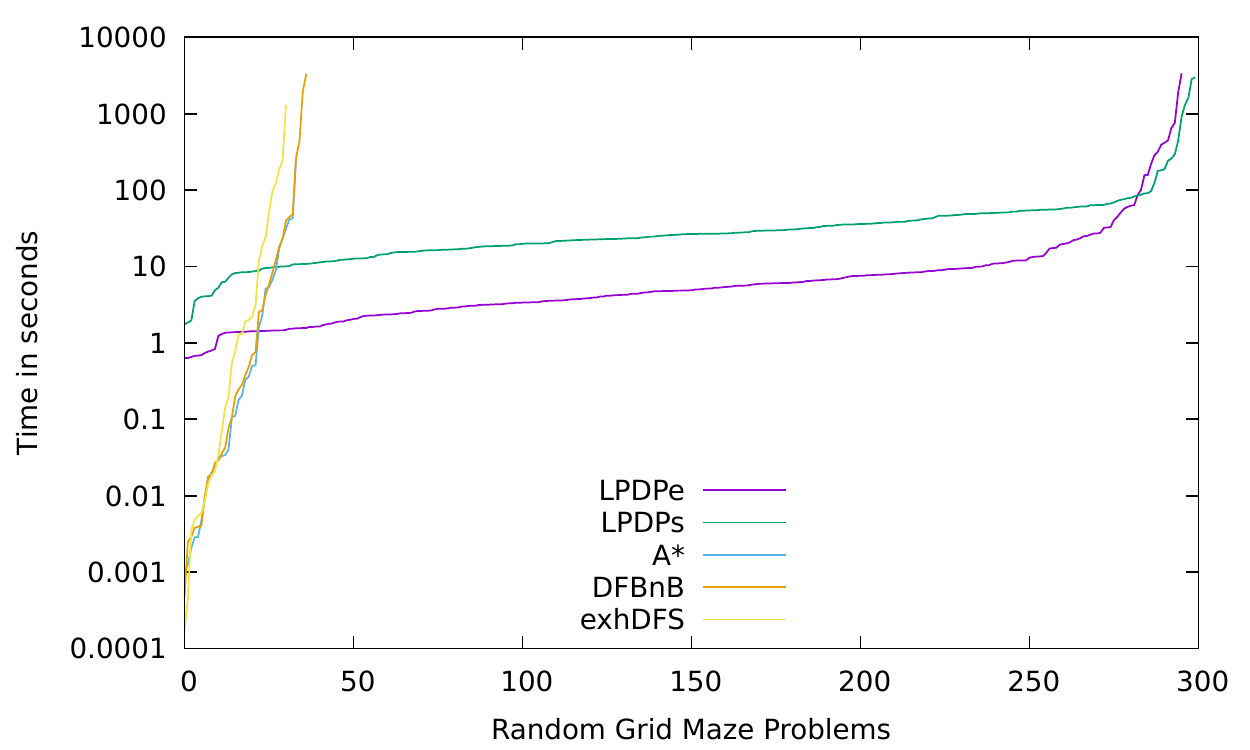}}
  \centerline{\includegraphics[width=\linewidth]{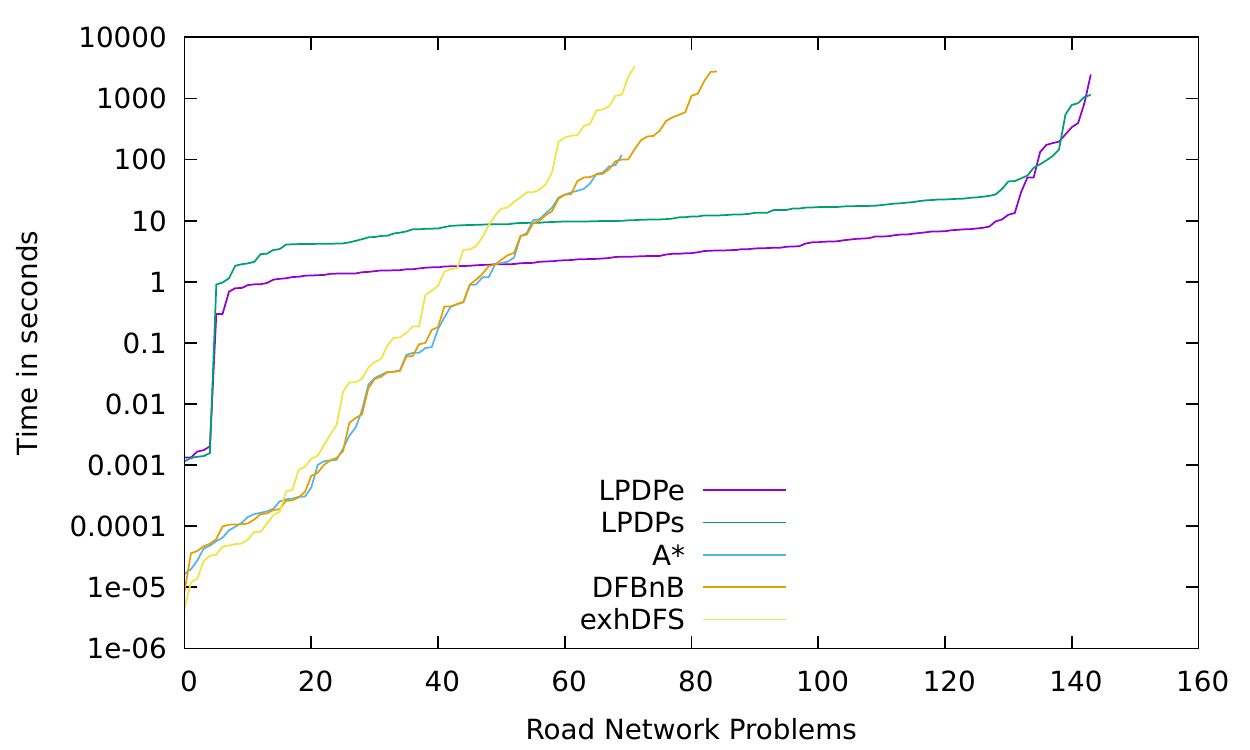}}
  \centerline{\includegraphics[width=\linewidth]{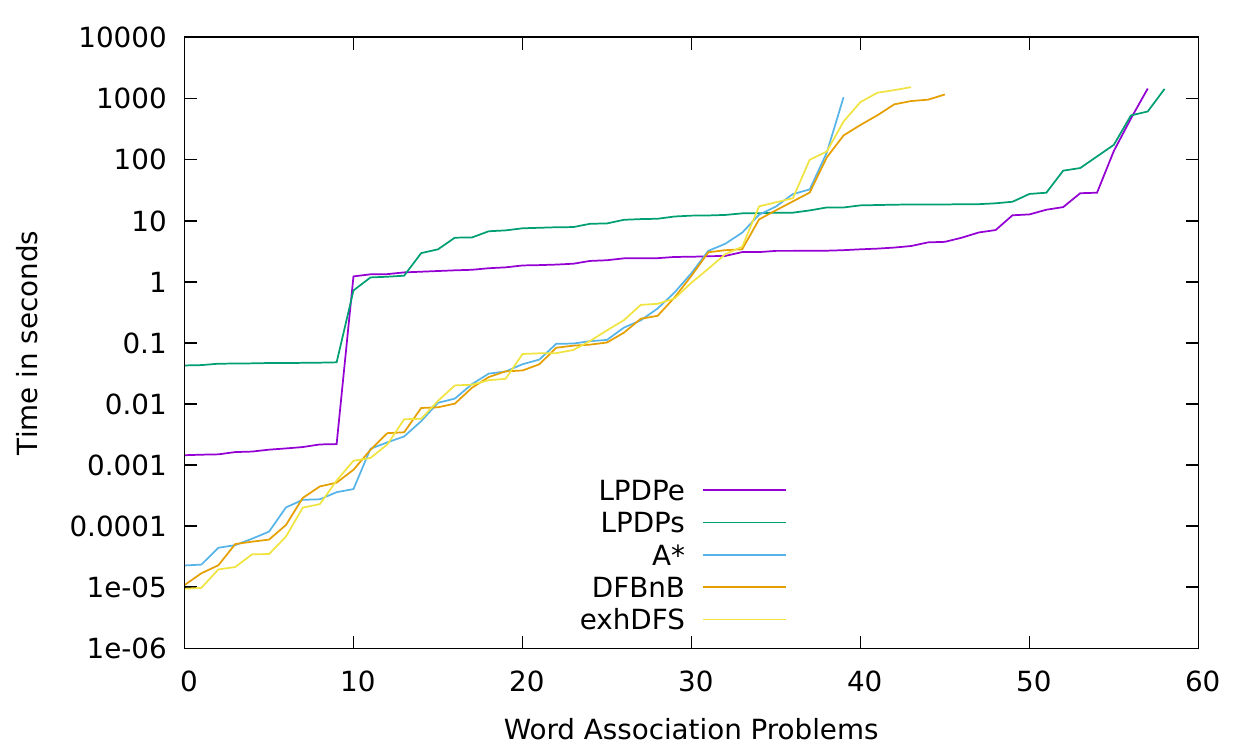}}
  \caption{Cactus plots for the three kinds of benchmark problems comparing previous algorithms to
  LPDP with three different partitioning configurations. Running times include time spent on
  partitioning for the LPDP variants.}
  \label{fig:cactuses}
  \vspace*{-.5cm}
\end{figure}

\begin{figure}
  \includegraphics[width=\linewidth]{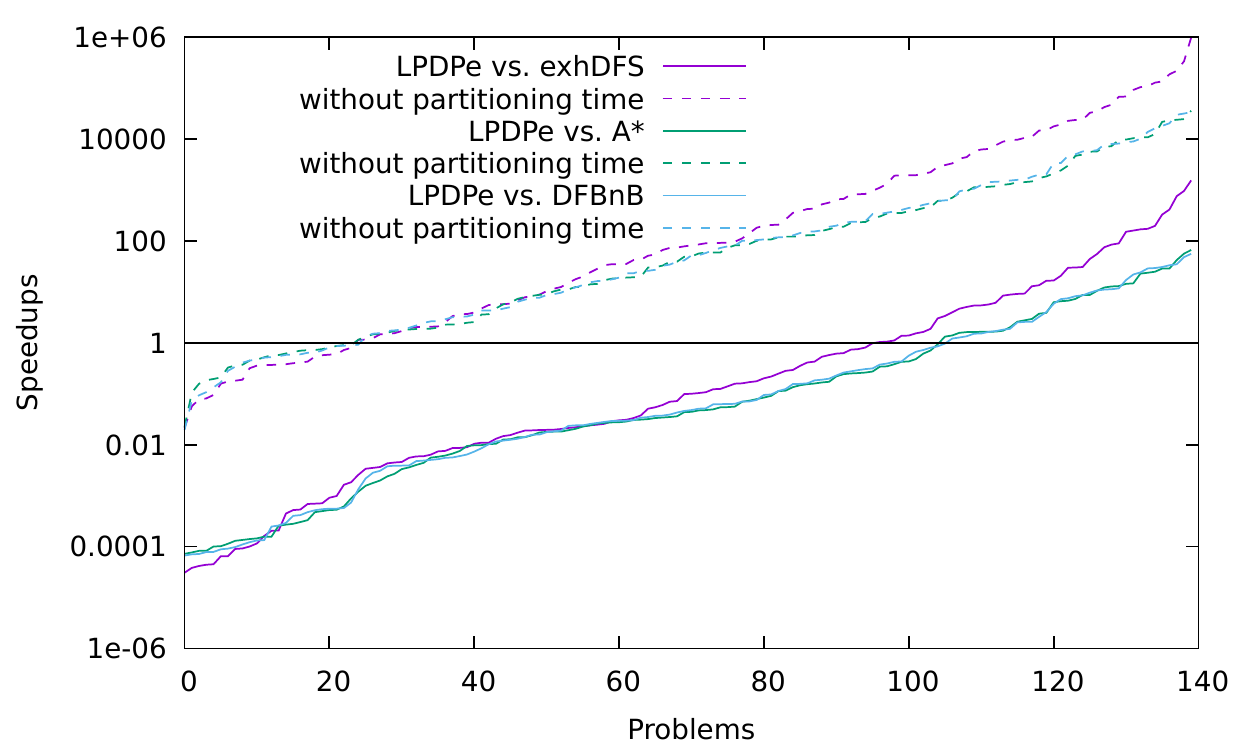}
  \caption{Speedup of LPDPe in relation to previous LP algorithms on 
  problems that were solved within the time limit by each of the tested algorithms. The dashed lines show the speedup if we only measure the execution time of the LPDP algorithm and ignore partitioning time.}
  \label{fig:speedup}
\end{figure}

\begin{figure}[t]
\begin{center}
  \includegraphics[width=\linewidth]{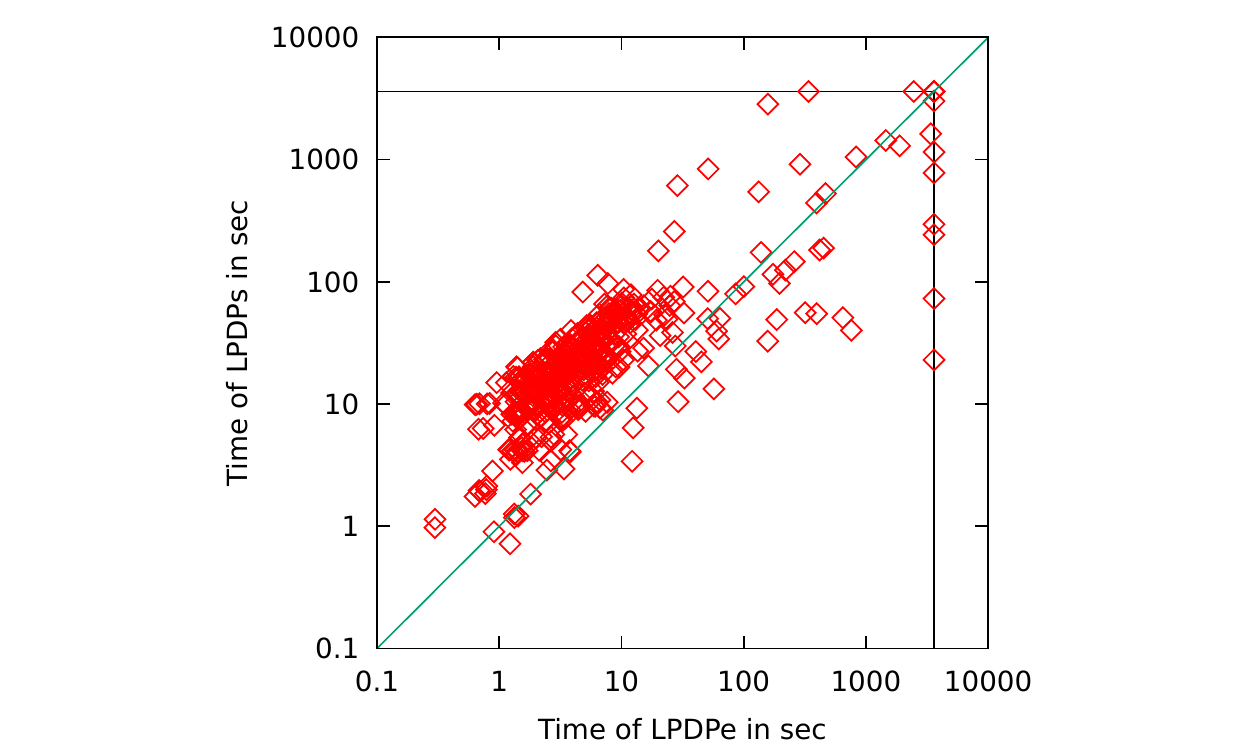}
  \end{center}
  \caption{A scatter plot comparing running times of LPDPe  and LPDPs  
  on the entire benchmark set.
  Points above (below) the green line represent instances where LPDPe (LPDPs) 
  was faster.
  Points on the right (top) blue line represent instances that were not solved within the time limit of
  one hour by LPDPe (LPDPs).}
  \label{fig:scatter}
\end{figure}

\begin{table}
\begin{center}
\begin{tabular}{l||rrrr} 
  \multirow{2}{*}{Solver} & \multicolumn{4}{c}{Number of Solved Instances}\\ 
  & Grids & Roads & Words & Total\\ 
 \hline
 A*              & 34 & 70 & 40 &  144\\ 
 DFBnB           & 37 & 85 & 46 &  168\\ 
 Exhaustive DFS & 31 & 72 & 44 &  147\\ 
 \hline
 LPDPe                  & 296 & 144 & 58 & 498 \\ 
 LPDPs & 300 & 144 & 59 & 503 \\ 
\end{tabular}
\end{center}
\caption{The number of instances solved within the time limit of one hour 
by the tested solver configurations for each collection of benchmark problems and in total.}
\label{tab:solved}
\end{table}

The first set of instances is generated by using mazes in~$N\times N$ grids of square cells with a given start and target cell. 
One can move to adjacent cells horizontally or vertically but only if the cell is not an obstacle.
The goal is to find the longest simple path from the start cell to the target cell. 
We represent the grids as graphs: for every free cell we insert a vertex and we add an edge with weight 1 between any two vertices, whose cell are horizontally or vertically adjacent to each other.
We generate the grids as \citet{stern}: the top left and bottom right cell are the start and target cell. Random cells of the grid are consecutively made into obstacles until a certain percentage $x\in\{30\%, 40\%\}$ of all cells is filled. Afterwards a path between the start and target is searched for to make sure that a solution of the longest~path~problem~exists.
The sizes of the used mazes range from 10x10 up to 120x120 with 300 instances in total.

The second and third set of instances are subgraphs of the road network of New York as well as subgraphs of the word association graph~\cite{webgraphWS1,webgraphWS2}, respectively. A subgraph is extracted as follows: we start a breadth-first search from a random vertex of the network and stop it when a certain number of vertices is reached. The vertices touched by the breadth-first search induce the instance. 
One of the touched vertices is randomly chosen as the target-vertex. The sizes of the road network
subgraphs are 2,4,...,300 vertices, i.e., 150 instances in total. As for the word association benchmark set,
we have ten instances for each of the six sizes (10,20,...,60 vertices) -- in total 60 instances.

\subsection{Experimental Results}
We now compare A*, DFBnB, and exhDFS presented by \citet{stern} to
our algorithm LPDP using two configurations. Our configurations differ in the amount of time that we spent on partitioning the input instance. We use either the eco/default configuration of KaFFPa (\emph{LPDPe}), which is a good trade off between solution quality and running time, or the strong configuration of KaFFPaE which aims at better partitions while investing more time for partitioning (\emph{LPDPs}). In the latter case, we set the amount of block imbalance to 10\%.
Note that LPDPs spends much more time in the graph partitioning phase of the algorithm than LDPDe. 
All results reporting running time in this paper include the time spent~for~partitioning.

Figures~\ref{fig:cactuses}--\ref{fig:scatter} and Table~\ref{tab:solved} summarize the results of our experiments.
It is apparent from the cactus plots in Figure~\ref{fig:cactuses} that both configurations of LPDP significantly outperform the previous algorithms for
each kind of tested benchmark except for very easy problems. These problems are typically solved under a few seconds by any of the
algorithms. In these cases, most of the time of our algorithm is spent in the partitioning phase. Moreover, our LPDP algorithms can solve significantly more problems, which can be seen in the cactus
plots as well as in Table~\ref{tab:solved}.

There are 140 problem instances that were solved by all solvers within the time limit.
In Figure~\ref{fig:speedup} we provide the speedup of LPDPe against the three original LP algorithms.
For most of these instances the speedup is actually below~1, but from our data we know that this
happens only for easy problems (solvable within a couple of seconds by each solver). The slowdown on these easy instances is due to
the overhead caused by partitioning. Therefore Figure~\ref{fig:speedup} also shows the speedups if we only count the execution time of the LPDP algorithm itself while ignoring the partitioning time. We see that LPDP itself quickly outperforms the other solvers. Speedups below 1 only happen for very easy problems. If we include the partitioning time, using LPDPe only eventually pays off as the other solvers reach their limit.
A similar plot for LPDPs is not included as it looks very similar. The only significant difference is that LPDPs takes even longer to outperform the other algorithms. This is expected as LPDPs intentionally spends more time partitioning the~graph.


The differences in running time are highest for the grid maze instances and lowest for word association graph problems.
We believe this is due to the structure of these graphs, in particular, how well they can be partitioned
to loosely connected subgraphs. Our algorithm excels on problems that can be successfully partitioned but
is competitive on all~kinds~of~graphs.

As of evaluating our algorithm with different partitioning configurations, we see that spending extra time
on partitioning to get better solutions pays off. In particular, LPDPs is able to solve more instances. Especially if the instance appears to be hard it is worth while to invest more time in partitioning. Additionally, this depends
on how well the graphs can be partitioned (highest for grid mazes, smallest for word association). 

Looking at the scatter plot in Figure~\ref{fig:scatter}, we can see that LPDPe is faster for most of the
instances but has significantly more unsolved instances. Nevertheless, there are 
two instances that are solved by LDPDe and not by LPDPs. 
This shows that spending more effort on the partitioning does not necessarily increase the number of solved instances.

\subsection{Parallel Speedups}
\label{parallelexperiment}

\begin{table*}[]
\centering
\begin{tabular}{|c|c|c|ccc|ccc|}
 \hline
\multirow{2}{*}{Threads} & \multirow{2}{*}{\begin{tabular}[c]{@{}c@{}}Parallel\\ Solved\end{tabular}} & \multirow{2}{*}{\begin{tabular}[c]{@{}c@{}}Both\\ Solved\end{tabular}} & \multicolumn{3}{c|}{Speedup All} & \multicolumn{3}{c|}{Speedup Big} \\
    & & & Avg. & Tot. & Med. & Avg. & Tot. & Med.  \\ \hline
2 & 618
 & 618
 & \multicolumn{1}{c|}{1.038} & \multicolumn{1}{c|}{1.360} & 1.031 & \multicolumn{1}{c|}{1.335} & \multicolumn{1}{c|}{1.363} & 1.331 \\
 \hline
4 & 624
 & 621
 & \multicolumn{1}{c|}{1.392} & \multicolumn{1}{c|}{2.623} & 1.224 & \multicolumn{1}{c|}{2.542} & \multicolumn{1}{c|}{2.638} & 2.564 \\
 \hline
8 & 627
 & 621
 & \multicolumn{1}{c|}{1.788} & \multicolumn{1}{c|}{4.833} & 1.189 & \multicolumn{1}{c|}{4.707} & \multicolumn{1}{c|}{4.913} & 4.720 \\
 \hline
16 & 628
 & 621
 & \multicolumn{1}{c|}{2.257} & \multicolumn{1}{c|}{8.287} & 1.127 & \multicolumn{1}{c|}{8.097} & \multicolumn{1}{c|}{8.569} & 8.208 \\
 \hline
32 & 629
 & 621
 & \multicolumn{1}{c|}{2.344} & \multicolumn{1}{c|}{10.714} & 0.987 & \multicolumn{1}{c|}{11.272} & \multicolumn{1}{c|}{11.553} & 11.519 \\
 \hline
64 & 633
 & 621
 & \multicolumn{1}{c|}{2.474} & \multicolumn{1}{c|}{11.691} & 0.889 & \multicolumn{1}{c|}{15.180} & \multicolumn{1}{c|}{13.512} & 14.665 \\
 \hline
\end{tabular}
\caption{Parallel runtime speedup table: the serial version of LPDP had a time limit of 64 minutes for each problem. All parallel versions had a time limit of 32 minutes. A solver with $n$ threads considers a problem ``big'' if the serial solver took more than $n \cdot  5$ seconds to solve it.}
\label{speedup-table-1}
\end{table*}

In order to measure the speedups achieved through parallelization, we ran the solver multiple times on each problem. We first ran the serial solver (1 thread) and then doubled the number of threads with each additional run. This was done until 64 threads, the maximum number of threads that the hardware can support, were reached. It is to mention that the computer only has 32 cores. Simultaneous multithreading is used to support 2 simultaneous threads per core. This should reduce the maximum achievable speedup as two threads have to share the resources of one core. The serial solver had 64 minutes to solve each problem. The parallel solvers only had a time limit of 32 minutes.

Table \ref{speedup-table-1} gives a first overview of the results. In the second column of the table we see the number of problems that were solved by each of the parallel solvers. As expected this number increases as we increase the number of threads. With the 64 thread solver 15 problems remain unsolved. 
The third column shows the number of problems that were also solved by the serial algorithm. With 2 threads we initially solve fewer problems than with one. This is because of the decreased time limit given to the parallel solvers. From now on we will only look at this subset of problems as only they can be used to calculate the speedups achieved through parallelization. The average, total and median speedups can be seen in column 4, 5 and 6. The total speedup of a solver is how much faster it solved all the problems compared to the serial solver. For 2 threads we see a total speedup of 1.360. Doubling the thread count results in an increase of the total speedup by 92.9\%{}, 84.3\%{}, 71.5\%{}, 29.3\%{} and 9.1\%{}. We see that the initial jump from 2 to 4 threads almost doubles the speedup. The speedup gain per jump then slightly decreases until 16 threads are reached. At this point the total speedup is roughly half the number of threads (8.287). Further increasing the number of threads has a smaller effect on the total speedup. 32 threads still result in a gain of 29.3\%{}. Especially the final jump from 32 to 64 threads with 9.1\%{} only does little. We end with a total speedup of 11.691.

When looking at the average speedup per problem we see that it stays below 2.5 for all solvers. This is vastly different from the total speedup. This indicates that many problems in our benchmark set are relatively easy to solve. The overhead of the parallelization makes up a large part of the runtime for these problems. This keeps the average speedup small. The total speedup on the other hand is dominated by a smaller number of difficult problems which make up a large part of the total runtime of the benchmark.

The same thing can be seen when looking at the median speedups. Initially there is almost no speedup. With 4 threads the median speedup increases, but then starts to drop again. The 32 and 64 thread solver even have a median speedup below 1. This means that they are slower than the single threaded solver for at least half of the problems. From the data we know that none of these are difficult problems. LPDP solves them so fast that they do not warrant the necessary runtime overhead of parallelization.

In order to filter out these problems that are too easy to benefit from parallelization we restrict ourselves to ``big'' problems. For a solver with $n > 1$ threads we call a problem big if the serial solver took more than $5 \cdot n$ seconds to solve it. So 10, 20, 40, 80, 160 and 320 seconds for 2, 4, 8, 16, 32 and 64 threads. The threshold for a big problem increases with the number of threads as a higher thread count only pays off for more difficult problems. This can be seen in the steady decrease in the median speedup from 2 threads onwards.

The average, total and median speedups for big problems can be seen in column 7, 8 and 9 of the table. 
The total speedups for big problems are higher overall. 
The percentage gains when doubling the number of threads also are higher: 93.5\%{}, 86.2\%{}, 74.4\%{}, 34.8\%{} and 17.0\%{}. 
Especially the last jump from 32 to 64 threads now increases the total speedup by 17.0\%{} compared to the 9.1\%{} before. 
The biggest difference can be seen for the average and median speedups. Now they are relatively similar to the total speedups. 
An interesting point is that for all numbers of threads except for 64 the average and median speedup is slightly lower than the total speedup. 
For 64 threads both are higher than the total speedup. 
This means that some of the easier big problems give us higher speedups than the more difficult ones.

\section{Conclusion}
\label{s:conclusion}
We presented an exact algorithm for the longest path (LP) problem in undirected graphs which is based on dynamic programming and graph partitioning.
Experiments show that our new algorithm is faster for nontrivial problems than the previous exact algorithms and can solve significantly more benchmark instances if a time limit per instance is given.

We also presented and evaluated a parallel version of the algorithm in a shared-memory setting. We observed
speedup for up to 64 solver threads. For future work we plan to develop a parallel version for
computer clusters using message passing protocols.

\subsubsection*{Acknowledgments} This work was partially supported by DFG grant SCHU 2567/1-2..

\bibliography{LP-FiegerK.37}

\begin{thebibliography}{}

\bibitem[\protect\citeauthoryear{Applegate \bgroup et al\mbox.\egroup
  }{1994}]{concorde}
Applegate, D.; Bixby, R.; Chv{\'a}tal, V.; and Cook, W.
\newblock 1994.
\newblock Finding cuts in the {TSP}: a preliminary report.
\newblock {\em The Mathematical Programming Symposium 1994, Ann Arbor,
  Michigan}.

\bibitem[\protect\citeauthoryear{Balyo, Fieger, and
  Schulz}{2019}]{kaLPHomePage}
Balyo, T.; Fieger, K.; and Schulz, C.
\newblock 2019.
\newblock {KaLP -- Karlsruhe Longest Paths Homepage}.
\newblock {\url{http://algo2.iti.kit.edu/kalp/index.html}}.

\bibitem[\protect\citeauthoryear{Boldi and Vigna}{2004}]{webgraphWS1}
Boldi, P., and Vigna, S.
\newblock 2004.
\newblock The {W}eb{G}raph framework {I}: {C}ompression techniques.
\newblock In {\em Proc. of the Thirteenth International World Wide Web
  Conference (WWW 2004)},  595--601.
\newblock Manhattan, USA: ACM Press.

\bibitem[\protect\citeauthoryear{Boldi \bgroup et al\mbox.\egroup
  }{2011}]{webgraphWS2}
Boldi, P.; Rosa, M.; Santini, M.; and Vigna, S.
\newblock 2011.
\newblock Layered label propagation: A multiresolution coordinate-free ordering
  for compressing social networks.
\newblock In Srinivasan, S.; Ramamritham, K.; Kumar, A.; Ravindra, M.~P.;
  Bertino, E.; and Kumar, R., eds., {\em Proceedings of the 20th international
  conference on World Wide Web},  587--596.
\newblock ACM Press.

\bibitem[\protect\citeauthoryear{Brucker}{1995}]{Brucker}
Brucker, P.
\newblock 1995.
\newblock {\em Scheduling Algorithms}.
\newblock Secaucus, NJ, USA: Springer-Verlag New York, Inc.

\bibitem[\protect\citeauthoryear{Garey and Johnson}{1979}]{NP}
Garey, M.~R., and Johnson, D.~S.
\newblock 1979.
\newblock {\em Computers and Intractability: A Guide to the Theory of
  NP-Completeness}.
\newblock New York, NY, USA: W. H. Freeman \& Co.

\bibitem[\protect\citeauthoryear{Hardgrave and
  Nemhauser}{1962}]{hardgrave1962relation}
Hardgrave, W., and Nemhauser, G.~L.
\newblock 1962.
\newblock On the relation between the traveling-salesman and the longest-path
  problems.
\newblock {\em Operations Research} 10(5):647--657.

\bibitem[\protect\citeauthoryear{Heule \bgroup et al\mbox.\egroup
  }{2011}]{heule}
Heule, M.~J.; Kullmann, O.; Wieringa, S.; and Biere, A.
\newblock 2011.
\newblock Cube and conquer: Guiding {CDCL} {SAT} solvers by lookaheads.
\newblock In {\em Haifa Verification Conference},  50--65.
\newblock Springer.

\bibitem[\protect\citeauthoryear{Knuth}{1973}]{involution}
Knuth, D.~E.
\newblock 1973.
\newblock {\em The Art of Computer Programming, Volume 3: Sorting and
  Searching}.
\newblock Addison Wesley Longman Publishing Co., Inc.

\bibitem[\protect\citeauthoryear{Lawler \bgroup et al\mbox.\egroup
  }{1985}]{lawler1985traveling}
Lawler, E.~L.; Lenstra, J.~K.; Kan, A.~R.; and Shmoys, D.~B.
\newblock 1985.
\newblock {\em The traveling salesman problem. A guided tour of combinatorial
  optimisation}.
\newblock John Wiley \& Sons.

\bibitem[\protect\citeauthoryear{Ozdal and Wong}{2006a}]{Circuit1}
Ozdal, M.~M., and Wong, M. D.~F.
\newblock 2006a.
\newblock Algorithmic study of single-layer bus routing for high-speed boards.
\newblock {\em IEEE Transactions on Computer-Aided Design of Integrated
  Circuits and Systems} 25(3):490--503.

\bibitem[\protect\citeauthoryear{Ozdal and Wong}{2006b}]{Circuit2}
Ozdal, M.~M., and Wong, M. D.~F.
\newblock 2006b.
\newblock A length-matching routing algorithm for high-performance printed
  circuit boards.
\newblock {\em IEEE Transactions on Computer-Aided Design of Integrated
  Circuits and Systems} 25(12):2784--2794.

\bibitem[\protect\citeauthoryear{Palombo \bgroup et al\mbox.\egroup
  }{2015}]{palombo2015solving}
Palombo, A.; Stern, R.; Puzis, R.; Felner, A.; Kiesel, S.; and Ruml, W.
\newblock 2015.
\newblock Solving the snake in the box problem with heuristic search: First
  results.
\newblock In {\em Eighth Annual Symposium on Combinatorial Search}.

\bibitem[\protect\citeauthoryear{Portugal and Rocha}{2010}]{Multirobot}
Portugal, D., and Rocha, R.
\newblock 2010.
\newblock {MSP} algorithm: Multi-robot patrolling based on territory allocation
  using balanced graph partitioning.
\newblock In {\em Proceedings of the 2010 ACM Symposium on Applied Computing},
  SAC '10,  1271--1276.
\newblock New York, NY, USA: ACM.

\bibitem[\protect\citeauthoryear{Reinelt}{1991}]{tsplib}
Reinelt, G.
\newblock 1991.
\newblock {TSPLIB} -- a traveling salesman problem library.
\newblock {\em ORSA journal on computing} 3(4):376--384.

\bibitem[\protect\citeauthoryear{Robison}{2011}]{DBLP:reference/parallel/Robison11}
Robison, A.~D.
\newblock 2011.
\newblock Intel{\textregistered} threading building blocks {(TBB)}.
\newblock In Padua, D.~A., ed., {\em Encyclopedia of Parallel Computing}.
  Springer.
\newblock  955--964.

\bibitem[\protect\citeauthoryear{Sanders and Schulz}{2013}]{KaHIP}
Sanders, P., and Schulz, C.
\newblock 2013.
\newblock {Think Locally, Act Globally: Highly Balanced Graph Partitioning}.
\newblock In {\em Proceedings of the 12th International Symposium on
  Experimental Algorithms (SEA'13)}, volume 7933 of {\em LNCS},  164--175.
\newblock Springer.

\bibitem[\protect\citeauthoryear{Stern \bgroup et al\mbox.\egroup
  }{2014}]{stern}
Stern, R.; Kiesel, S.; Puzis, R.; Feller, A.; and Ruml, W.
\newblock 2014.
\newblock Max is more than min: Solving maximization problems with heuristic
  search.
\newblock In {\em Proceedings of the Seventh Annual Symposium on Combinatorial
  Search (SoCS 2014)}.

\bibitem[\protect\citeauthoryear{Wong, Lau, and King}{2005}]{Wong}
Wong, W.~Y.; Lau, T.~P.; and King, I.
\newblock 2005.
\newblock Information retrieval in {P2P} networks using genetic algorithm.
\newblock In {\em Special Interest Tracks and Posters of the 14th International
  Conference on World Wide Web}, WWW '05,  922--923.
\newblock New York, NY, USA: ACM.

\end{thebibliography}
\bibliographystyle{aaai}
\end{document}